\def\IEEEsubmission{1}
\def\isConference{1}

\def\figuresize{3.3in}

\def\complexNumbers{\mathbb{C}}

\def\functionSpace[#1]{\mathcal{F}(#1)}

\def\constante{{\rm e}}
\def\constantj{{\rm j}}
\def\expectationOperator[#1][#2]{\mathbb{E}_{#2}[#1]}
\def\varianceOperator[#1]{\mathrm{Var}\left(#1\right)}
\def\uniformDistribution[#1][#2]{{\mathcal{U}_{[#1,#2]}}}
\def\traceOperator[#1]{{\mathrm{tr}}\{#1\}}
\def\identityMatrix[#1]{\mathrm{\textbf{I}}_{#1}}
\def\zeroVector[#1]{{ {{\mathbf{0}}}}_{#1}}
\def\oneVector[#1]{{ {\mathbf{1}}}_{#1}}
\def\exponentialIntegral[#1]{\mathrm{Ei}(#1)}
\def\functionArbitrary[#1]{f_{#1}}
\def\functionArbitraryEstimate[#1]{\hat{f}_{#1}}
\def\indicatorFunction[#1]{\mathbb{I}\left[{#1}\right]}

\def\clamp[#1][#2]{\text{clamp}_{#2}\left(#1\right)}
\def\signNormal[#1]{\text{sign}\left(#1\right)}
\def\diagOperation[#1]{\text{diag}\left\{#1\right\}}

\def\probability[#1]{\mathrm{Pr}({#1})}
\def\complexGaussian[#1][#2]{\mathcal{CN}({#1,#2})}
\def\gaussian[#1][#2]{\mathcal{N}({#1,#2})}
\def\rayleigh[#1]{\mathrm{Rayleigh}({#1})}

\def\normalPDF[#1]{\phi\left(#1\right)}
\def\normalCDF[#1]{\Phi\left(#1\right)}

\def\CDF[#1][#2][#3]{F_{#1}^{#2}\left({#3}\right)}
\def\PDF[#1][#2][#3]{f_{#1}^{#2}\left({#3}\right)}

\def\complexGaussian[#1][#2]{\mathcal{CN}({#1,#2})}

\def\indexED{k}

\def\desiredPhase{\theta_{\rm desired}}
\def\errPhase[#1]{\theta_{#1}^{\text{err}}}
\def\errPhaseCFO[#1]{\theta_{#1}^{\text{cfo}}}
\def\errPhaseMobility[#1]{\theta_{#1}^{\text{mob}}}
\def\errPhaseNoise[#1]{\theta^\text{noise}_{#1}}
\def\phaseSymbol[#1]{{\hat{\theta}_{#1}}}
\def\phaseSymbolWithoutNoise[#1]{\theta_{#1}}
\def\pilotSymbol[#1]{p_{#1}}

\def\timeSymbol{t}
\def\timeSymbolwithIndex[#1][#2]{t_{#1#2}}
\def\instantStepOne[#1]{T_{#1}^{(1)}}
\def\instantStepTwo[#1]{T_{#1}^{(2)}}
\def\instantStepThree[#1]{T_{#1}^{(3)}}
\def\instantStepFour[#1]{T_{#1}^{(4)}}

\def\phaseEstimateStepOne[#1]{{\hat{\theta}_{#1}^{(1)}}}
\def\phaseEstimateStepTwo[#1]{{\hat{\theta}_{#1}^{(2)}}}
\def\phaseEstimateStepThree[#1]{{\hat{\theta}_{#1}^{(3)}}}
\def\phaseEstimateStepFour[#1]{{\hat{\theta}_{#1}^{(4)}}}

\def\noiseSymbol[#1]{\omega_{#1}}
\def\noiseAmp[#1]{n_{#1}}
\def\noisePha[#1]{\phi_{#1}}
\def\noisePhase[#1]{\psi_{#1}}
\def\transmitPower[#1]{P_{#1}}
\def\noiseVariance{\sigma_{\rm n}^2}
\def\noiseVarianceED{\sigma_{\rm noise@UE}^2}
\def\noiseVarianceES{\sigma_{\rm noise@BS}^2}

\def\SNRatED{{\tt SNR}_{\rm @UE}}
\def\SNRatES{{\tt SNR}_{\rm @BS}}

\def\noiseSymbolStepOne[#1]{\omega_{#1}^{(1)}}
\def\noiseAmpStepOne[#1]{n_{#1}^{(1)}}
\def\noiseAngleStepOne[#1]{\phi_{#1}^{(1)}}
\def\noisePhaseAdditiveStepOne[#1]{\psi_{#1}^{(1)}}

\def\noiseSymbolStepTwo[#1]{\omega_{#1}^{(2)}}
\def\noiseAmpStepTwo[#1]{n_{#1}^{(2)}}
\def\noiseAngleStepTwo[#1]{\phi_{#1}^{(2)}}
\def\noisePhaseAdditiveStepTwo[#1]{\psi_{#1}^{(2)}}

\def\noiseSymbolStepThree[#1]{\omega_{#1}^{(3)}}
\def\noiseAmpStepThree[#1]{n_{#1}^{(3)}}
\def\noiseAngleStepThree[#1]{\phi_{#1}^{(3)}}
\def\noisePhaseAdditiveStepThree[#1]{\psi_{#1}^{(3)}}

\def\noiseSymbolStepFour[#1]{\omega_{#1}^{(\text{4})}}
\def\noiseAmpStepFour[#1]{n_{#1}^{(\text{4})}}
\def\noiseAngleSteFour[#1]{\phi_{#1}^{(\text{4})}}
\def\noisePhaseAdditiveStepFour[#1]{\psi_{#1}^{(\text{4})}}

\def\symbolEle[#1]{{{d}}_{#1}}
\def\carrierFrequency{f_{\rm c}}

\def\numberOfEdgeDevices{K}

\def\waveLength{\lambda}

\def\transmittedSymbol[#1][#2]{{x}_{#1}}
\def\receivedSymbol{r}

\def\transmittedSymbolUL[#1][#2]{{x}_{#1}}
\def\receivedSymbolUL[#1][#2]{{r}_{#1,#2}}
\def\transmittedSymbolDL[#1][#2]{{x}_{#1}}
\def\receivedSymbolDL[#1][#2]{{y}_{#1,#2}}
\def\superposedSymbol[#1]{{s}_{#1}}

\def\channelEle[#1][#2]{h^{#1}_{#2}}

\def\channelAmp[#1]{a_{#1}}
\def\channelPhase[#1]{\theta_{#1}}
\def\channelAmpDL[#1]{a_{#1}^{\text{d}}}
\def\channelPhaseDL[#1]{\phi_{#1}^{\text{d}}}
\def\channelAmpUL[#1]{a_{#1}^{\text{u}}}
\def\channelPhaseUL[#1]{\phi_{#1}^{\text{u}}}

\def\channelPhaseDLInitial[#1]{\phi_{#1,0}^{\text{d}}}
\def\channelPhaseULInitial[#1]{\phi_{#1,0}^{\text{u}}}

\def\channelAmpNewDL[#1]{a'_{\text{d},#1}}
\def\channelPhaseNewDL[#1]{\phi'_{\text{d},#1}}
\def\channelAmpNewUL[#1]{a'_{\text{u},#1}}
\def\channelPhaseNewUL[#1]{\phi'_{\text{u},#1}}

\def\maxVelocity{v}
\def\velocity[#1]{v_{#1}}
\def\speedOfLigth{c}

\def\CFO[#1]{\Delta f_{#1}}
\def\PO[#1]{\delta_{#1}}

\def\angleOfArrival[#1]{\alpha_{#1}}
\def\numberOfOACSymbols{M}

\def\functionArbitrary[#1]{f_{#1}}
\def\preProcessingFunction[#1][#2]{\psi_{#1}{\left(#2\right)}}
\def\postProcessingFunction[#1][#2]{\varphi_{#1}{\left(#2\right)}}

\def\guardTimeDownlink{g_\text{d}}
\def\guardTimeUplink{g_\text{u}}
\def\packetDuration{T_\text{p}}

\def\varianceCFO{\sigma_{\text{cfo}}^{2}}
\def\noisePhaseAgg[#1]{\epsilon_{#1}}

\def\indexOACsymbol{m}
\def\symbolDurationOFDMTotal{T_{\text{s}}}
\def\symbolDurationOFDM{T_{\text{ofdm}}}
\def\cpDuration{T_{\text{cp}}}

\def\RMSE[#1]{\textrm{RMSE}_{#1}}

\def\velocityMean{\bar{v}}

\def\numberOfBitsForQuantization{Q}
\def\spectralEfficiency{r_{\text{eff}}}

\def\stdPhaseError[#1][#2]{\sigma^{\text{err}}_{#1,#2}}

\if\isConference1
	\documentclass[conference]{IEEEtran}
\else
	\if\IEEEsubmission1
	\documentclass[journal,12pt,onecolumn,draftclsnofoot]{IEEEtran}
	\else
	\documentclass[journal]{IEEEtran}
	\fi
\fi

\usepackage{multicol}
\usepackage{lipsum}
\usepackage{mathtools}
\usepackage{amsthm}
\usepackage{acronym}
\usepackage[bookmarksopen=true]{hyperref}
\usepackage{stfloats}
\usepackage{amsfonts}
\usepackage{acronym}
\usepackage[noadjust]{cite}
\usepackage{multirow}
\usepackage{bm}
\usepackage{amsmath,amssymb}
\usepackage{graphicx}
\usepackage{epstopdf}
\usepackage[table,xcdraw]{xcolor}
\usepackage[caption=false,font=footnotesize]{subfig}
\usepackage[geometry]{ifsym}
\usepackage{array}
\usepackage[utf8]{inputenc}
\usepackage[T1]{fontenc}
\usepackage{pifont}
\usepackage{tabularray}
\usepackage{lipsum}
\def\BibTeX{{\rm B\kern-.05em{\sc i\kern-.025em b}\kern-.08em
		T\kern-.1667em\lower.7ex\hbox{E}\kern-.125emX}}

\newcommand\mydots{\hbox to 1em{.\hss.\hss.}}

\makeatletter
\newif\ifAC@uppercase@first%
\def\Aclp#1{\AC@uppercase@firsttrue\aclp{#1}\AC@uppercase@firstfalse}%
\def\AC@aclp#1{%
	\ifcsname fn@#1@PL\endcsname%
	\ifAC@uppercase@first%
	\expandafter\expandafter\expandafter\MakeUppercase\csname fn@#1@PL\endcsname%
	\else%
	\csname fn@#1@PL\endcsname%
	\fi%
	\else%
	\AC@acl{#1}s%
	\fi%
}%
\def\Acp#1{\AC@uppercase@firsttrue\acp{#1}\AC@uppercase@firstfalse}%
\def\AC@acp#1{%
	\ifcsname fn@#1@PL\endcsname%
	\ifAC@uppercase@first%
	\expandafter\expandafter\expandafter\MakeUppercase\csname fn@#1@PL\endcsname%
	\else%
	\csname fn@#1@PL\endcsname%
	\fi%
	\else%
	\AC@ac{#1}s%
	\fi%
}%
\def\Acfp#1{\AC@uppercase@firsttrue\acfp{#1}\AC@uppercase@firstfalse}%
\def\AC@acfp#1{%
	\ifcsname fn@#1@PL\endcsname%
	\ifAC@uppercase@first%
	\expandafter\expandafter\expandafter\MakeUppercase\csname fn@#1@PL\endcsname%
	\else%
	\csname fn@#1@PL\endcsname%
	\fi%
	\else%
	\AC@acf{#1}s%
	\fi%
}%
\def\Acsp#1{\AC@uppercase@firsttrue\acsp{#1}\AC@uppercase@firstfalse}%
\def\AC@acsp#1{%
	\ifcsname fn@#1@PL\endcsname%
	\ifAC@uppercase@first%
	\expandafter\expandafter\expandafter\MakeUppercase\csname fn@#1@PL\endcsname%
	\else%
	\csname fn@#1@PL\endcsname%
	\fi%
	\else%
	\AC@acs{#1}s%
	\fi%
}%
\edef\AC@uppercase@write{\string\ifAC@uppercase@first\string\expandafter\string\MakeUppercase\string\fi\space}%
\def\AC@acrodef#1[#2]#3{%
	\@bsphack%
	\protected@write\@auxout{}{%
		\string\newacro{#1}[#2]{\AC@uppercase@write #3}%
	}\@esphack%
}%
\def\Acl#1{\AC@uppercase@firsttrue\acl{#1}\AC@uppercase@firstfalse}
\def\Acf#1{\AC@uppercase@firsttrue\acf{#1}\AC@uppercase@firstfalse}
\def\Ac#1{\AC@uppercase@firsttrue\ac{#1}\AC@uppercase@firstfalse}
\def\Acs#1{\AC@uppercase@firsttrue\acs{#1}\AC@uppercase@firstfalse}

\acrodef{WSN}{wireless sensor network}
\acrodef{USRP}{universal software radio peripheral}
\acrodef{SN}{sensor node}
\acrodef{FC}{fusion center}
\acrodef{MAC}{multiple-access channel}
\acrodef{FL}{federated learning}
\acrodef{ED}{edge device}
\acrodef{CS}{compressed sensing}
\acrodef{ES}[BS]{base station}
\acrodef{DCN}{data center network}
\acrodef{RIS}{reconfigurable intelligent surfaces}
\acrodef{IMC}{in-memory computing}
\acrodef{FPGA}{field-programmable gate array}
\acrodef{SDR}{software-defined radio}
\acrodef{PS}{processing system}
\acrodef{SS}{soft synchronization}
\acrodef{IQ}{in-phase/quadrature}
\acrodef{IP}{intellectual property}
\acrodef{DMA}{direct-memory access}
\acrodef{RAM}{random access memory}
\acrodef{CC}{companion computer}
\acrodef{FEE}{function estimation error}
\acrodef{MSK}{minimum-shift keying}
\acrodef{TDMA}{time-domain multiple access}
\acrodef{PLNC}{physical-layer network coding}
\acrodef{UAV}{unmanned aerial vehicle}
\acrodef{LoRa}{Long-Range}
\acrodef{DC}{direct-current}
\acrodef{DAC}{digital-to-analog converter}
\acrodef{ADC}{anlog-to-digital converter}
\acrodef{CS}{complementary sequence}
\acrodef{GCP}{Golay complementary pair}
\acrodef{ANF}{algebraic normal form}
\acrodef{AACF}{aperiodic auto-correlation function}
\acrodef{RM}{Reed-Muller}
\acrodef{MOCZ}{modulation on conjugate-reciprocal zeros}
\acrodef{BMOCZ}{binary modulation on conjugate-reciprocal zeros}
\acrodef{dizet}[DiZeT]{direct zero-testing}

\acrodef{PCP}[PCP]{phase-coded pilot}

\acrodef{PUCCH}{physical uplink control channel}
\acrodef{PRACH}{physical random access channel}

\acrodef{OBO}{output-power back-off}
\acrodef{ACLR}{adjacent-channel-leakage ratio}

\acrodef{LDPC}{low-density parity check}

\acrodef{PDF}{probability density function}
\acrodef{CDF}{cumulative distribution function}

\acrodef{TBMA}{type-based multiple access}

\acrodef{MSFE}{mean-squared function error}
\acrodef{FEE}{function-estimation error}
\acrodef{CER}{computation error rate}
\acrodef{BCER}{block-computation error rate}
\acrodef{CFO}{carrier frequency offset}
\acrodef{TO}{time offset}
\acrodef{PO}{phase offset}
\acrodef{RSSI}{received signal strength  information}

\acrodef{STLC}{space-time line code}
\acrodef{CCI}{co-channel interference}
\acrodef{CSIT}[CSIT]{\ac{CSI} at the transmitter}
\acrodef{CSIR}[CSIR]{\ac{CSI} at the receiver}
\acrodef{MIMO}{multiple-input-multiple-output}
\acrodef{PC}{phase correction}
\acrodef{ZF}{zero-forcing}
\acrodef{ANOVA}{analysis of variance}

\acrodef{PCA}{principal component analysis}
\acrodef{TIG}{Technical Interest Group}

\acrodef{FSK}{frequency-shift keying}
\acrodef{PPM}{pulse-position modulation}
\acrodef{PAM}{pulse-amplitude modulation}

\acrodef{MRC}{maximum-ratio combining}
\acrodef{HP}{hard-coded participation}
\acrodef{HPA}{hard-coded participation with absentees}
\acrodef{SP}{soft-coded participation}
\acrodef{FSK-MV}{\ac{FSK}-based \ac{MV}}
\acrodef{RF}{radio-frequency}
\acrodef{MF}{matched filter}
\acrodef{PPM}{pulse-position modulation}
\acrodef{CSK}{chirp-shift keying}
\acrodef{PPM-MV}[PPM-MV]{\ac{PPM}-based \ac{MV}}
\acrodef{DFT-s-OFDM}{discrete Fourier transform-spread orthogonal frequency division multiplexing}
\acrodef{SC}{single-carrier}
\acrodef{SGD}{stochastic gradient descent}
\acrodef{signSGD}{sign stochastic gradient descent}

\acrodef{SL}{split learning}
\acrodef{SNR}{signal-to-noise ratio}
\acrodef{RMSE}{root-mean-square error}
\acrodef{OFDM}{orthogonal frequency division multiplexing}
\acrodef{DFT}{discrete Fourier transform}
\acrodef{PSK}{phase-shift keying}
\acrodef{QAM}{quadrature amplitude modulation}
\acrodef{QPSK}{quadrature phase-shift keying}
\acrodef{PMEPR}{peak-to-mean envelope power ratio}
\acrodef{BER}{bit-error ratio}
\acrodef{SNR}{signal-to-noise ratio}
\acrodef{PSD}{power spectral density}
\acrodef{SE}{spectral efficiency}
\acrodef{CP}{cyclic prefix}
\acrodef{AWGN}{additive white Gaussian noise}
\acrodef{CFR}{channel frequency response}
\acrodef{CIR}{channel impulse response}
\acrodef{MMSE}{minimum mean-squared error}
\acrodef{LMMSE}{linear minimum mean-squared error}
\acrodef{BPSK}{binary phase shift keying}
\acrodef{QPSK}{quadrature phase shift keying}
\acrodef{BLER}{block-error rate}
\acrodef{ML}{machine learning}
\acrodef{MaxLike}{maximum likelihood}
\acrodef{PHY}{physical layer}
\acrodef{PA}{power amplifier}
\acrodef{UE}{user equipment}
\acrodef{BS}{base station}
\acrodef{IDFT}{inverse discrete Fourier transform}
\acrodef{DoF}{degrees-of-freedom}
\acrodef{IoT}{Internet-of-Things}
\acrodef{FDE}{frequency-domain equalization}
\acrodef{RF}{radio-frequency}
\acrodef{IM}{index modulation}
\acrodef{MF}{matched filter}
\acrodef{PPM}{pulse-position modulation}

\acrodef{MSE}{mean-squared error}
\acrodef{MRT}{maximum-ratio transmission}
\acrodef{ERC}{equal-ratio combining}
\acrodef{BAA}{broadband analog aggregation}
\acrodef{OBDA}{one-bit broadband digital aggregation}
\acrodef{FEEL}{federated edge learning}
\acrodef{FL}{federated learning}
\acrodef{UL}{uplink}
\acrodef{DL}{downlink}
\acrodef{OAC}{over-the-air computation}
\acrodef{TCI}{truncated-channel inversion}
\acrodef{MV}{majority vote}
\acrodef{CNN}{convolution neural network}
\acrodef{ReLU}{rectified-linear unit}
\acrodef{CSI}{channel state information}
\acrodef{PAPR}{peak-to-average power ratio}
\acrodef{SC}{single-carrier}
\acrodef{iid}[IID]{independent and identically distributed}
\acrodef{RMS}{root-mean-square}
\acrodef{4G}{Fourth Generation}
\acrodef{5G}{Fifth Generation}
\acrodef{NR}{New Radio}
\acrodef{LTE}{Long-Term Evolution}
\acrodef{OFDMA}{orthogonal frequency division multiple access}
\acrodef{ICI}{inter-carrier interference}
\acrodef{HARQ}{hybrid automatic repeat request}
\acrodef{D2D}{Device-to-Device}
\acrodef{NOMA}{non-orthogonal multiple access}
\acrodef{OMA}{orthogonal multiple access}

\acrodef{IMT}{International Mobile Telecommunications}
\acrodef{ITU}{International Telecommunication Union}

\acrodef{PDP}{power-delay profile}
\acrodef{TBMA}{type-based multiple access}

\IEEEoverridecommandlockouts
\begin{document}

\if\isConference0
	\title{
		{On the Feasibility of Distributed Phase Synchronization for Coherent Signal Superposition}\\
		\thanks{Alphan~\c{S}ahin is with the Electrical  Engineering Department,
			University of South Carolina, Columbia, SC, USA. E-mail: asahin@mailbox.sc.edu}	
		\author{Alphan~\c{S}ahin,~\IEEEmembership{Member,~IEEE}} 
	}
\else
	\title{On the Feasibility of Distributed Phase Synchronization for Coherent Signal Superposition}
	
    \author{Alphan \c{S}ahin\\
	Department of Electrical Engineering, University South Carolina, Columbia, SC, USA\\
	Email: asahin@mailbox.sc.edu 
					\thanks{This work has been supported by the National Science Foundation through the award CNS-2438837.}
}
\fi
\maketitle

\begin{abstract}
In this study, we analyze the feasibility of distributed phase synchronization for coherent signal superposition, a fundamental enabler for paradigms like coherent \ac{OAC}, distributed beamforming, and interference alignment, under mobility and hardware impairments.  With the focus on coherent \ac{OAC}, we introduce \acp{PCP},  a strategy where the radios communicate with each other to eliminate the round-trip phase change in the \ac{UL} and \ac{DL} to align the phase of the received symbol at a desired angle. In this study, considering a \ac{CFO}-resilient multi-user procedure, we derive the statistics of the phase deviations to assess how fast the phase coherency degrades. Our results show that residual \ac{CFO} is a major factor determining the duration of phase coherency, in addition to the non-negligible effects of mobility and the number of nodes in the network. We also provide a proof-of-concept demonstration for coherent signal superposition by using off-the-shelf radios to demonstrate the feasibility of PCPs in practice.
\end{abstract}
\section{Introduction}
\acresetall

Traditionally, communication and computation are viewed as separate tasks. This approach has been very effective from the engineering perspective, as isolated optimizations can be performed. However, for many computation-oriented applications, the ultimate goal is  to compute some mathematical functions (e.g., arithmetic mean, maximum, minimum) of the local information distributed at the different devices rather than the local information itself. In such scenarios, information theoretical results \cite{Jeon_2014} show that harnessing the interference for computation, i.e., \textit{\ac{OAC}}, can provide a significantly higher computation rate than separating communication and computation. 
The distinct feature of \ac{OAC} is that it does not acquire the data from each data-generating nodes via typical orthogonal multiple access methods. Instead, all the nodes transmit simultaneously, where the signal superposition in the channel leads to the desired computation result at the receiver; see the surveys in \cite{sahinSurvey2023,Zhibin_2022oac,pérezneira2024waveformscomputingair}. Thus, OAC reduces the latency scaled by the number of nodes. Hence, it is a  disruptive concept to the traditional way of handling computation and communication independently.

Although promising applications of OAC, such as wireless \ac{FL} \cite{Xiaowen_2022}, distributed control systems \cite{Cai_2018}, and wireless data centers \cite{Xiugang_2016}, have been proposed in the literature, to this date, \ac{OAC} has not been used in any communication standard or a commercial system. This is partially because  realizing a reliable \ac{OAC} in practice is  challenging due to the fading in wireless channels and hardware impairments such as \ac{CFO}, \ac{PO}, synchronization errors, and imperfect power control. In particular, to achieve \textit{coherent} \ac{OAC}, the radios must precode their transmissions so that the parameters add up constructively in the complex plane for coherent superposition. Otherwise, the phase of the signals arriving at the receiver will not be the same, destroying the desired coherent aggregation. Thus, the main bottleneck in coherent OAC is to estimate and design the precoder that counteracts the fading channels and hardware impairments, such as \ac{CFO} and \ac{PO} \cite{Guo_2021,Lizhao_2024tmc}. The same issue also arises in several important paradigms, such as interference alignment \cite{Moghadam2014}, distributed beamforming \cite{Mudumbai_2007distBeaming, Mghabghab_2021}, and physical-layer network coding \cite{Sachin_2008}, as they also rely on phase synchronization. Hence, effective solutions addressing the inevitable imperfections are needed to enable these important paradigms in the next-generation networks.

Given the enabling and fundamental nature of distributed phase synchronization, this paper analyzes its feasibility under hardware impairments and mobility. We introduce \textit{\acp{PCP}}, a method that allows the radios to eliminate the round-trip phase change in \ac{UL} and \ac{DL} channels, and discuss a \ac{CFO}-resilient multi-user procedure based on \acp{PCP} for OAC. The proposed approach does not rely on channel reciprocity, and it can be implemented in practice with off-the-shelf \acp{SDR}. In this work, in particular, we derive the statistics of the phase deviations for a given residual \ac{CFO} and mobility statistics to provide insights into the limits of phase synchronization for OAC, which can also be useful for other paradigms requiring distributed phase synchronization.

{\em Notation:} The set of complex numbers is $\complexNumbers$.
$\expectationOperator[\cdot][]$ is the expectation of its  argument. 
The zero-mean circularly symmetric complex Gaussian distribution with the variance $\sigma^2$, the uniform distribution with the support between $a$ and $b$, and the Rayleigh distribution with the scale $\sigma$ are $\complexGaussian[\sigma^2][0]$, $\uniformDistribution[a][b]$, and  $\rayleigh[\sigma]$, respectively.
The \ac{CDF} of the standard normal distribution is $\normalCDF[\cdot]$. 

\section{System Model}
\label{sec:system}

To shed light on how hardware impairment and fading channels affect \ac{OAC}, consider a scenario with  $\numberOfEdgeDevices$ nodes and one \ac{BS}, where all the nodes are triggered to transmit simultaneously. Let $\symbolEle[\indexED,\indexOACsymbol]\in\complexNumbers$ denote the $\indexOACsymbol$th parameter at the $\indexED$th node for $\indexOACsymbol\in\{0,1,\mydots,\numberOfOACSymbols-1\}$ and $\indexED\in\{{1,\mydots,\numberOfEdgeDevices}\}$, where $\numberOfOACSymbols$ is the number of parameters. Suppose that the goal of the \ac{BS} is to compute $\sum_{\indexED=1}^{\numberOfEdgeDevices}\symbolEle[\indexED,\indexOACsymbol]$, $\forall\indexOACsymbol$, via signal superposition on the complex plane. 

\begin{figure}
	\centering
	\includegraphics[width = \figuresize]{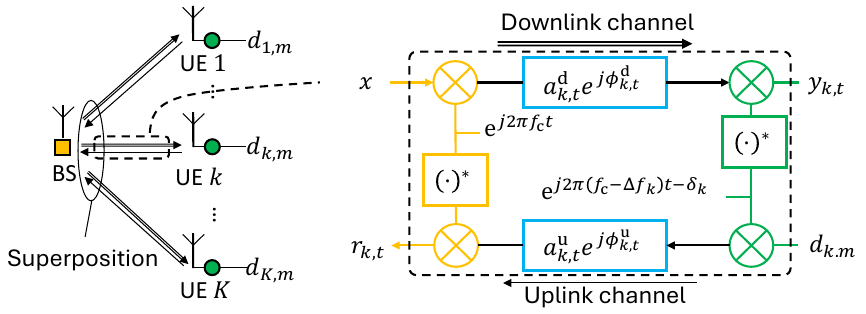}
	\caption{The channels  in UL and DL including environment, CFO, and PO.}
	\label{fig:systemModelHW}
	\vspace{-3mm}
\end{figure}
Consider the channel model shown in \figurename~\ref{fig:systemModelHW}, where  $\channelAmpUL[\indexED,\timeSymbol]\constante^{\channelPhaseUL[\indexED,\timeSymbol]}$ is the \ac{UL} channel coefficient,  $\channelAmpDL[\indexED,\timeSymbol]\constante^{\channelPhaseDL[\indexED,\timeSymbol]}$ is the \ac{DL} channel coefficient, and $\CFO[\indexED]$ and $\PO[\indexED]$ are the \textit{residual} \ac{CFO} and the \ac{PO} with respect to the local oscillator of the \ac{BS} with the carrier frequency $\carrierFrequency$, respectively. We assume that the same oscillator is used at the transmit and receive chains. Without any reciprocity assumption,
we model $\channelPhaseDL[\indexED,\timeSymbol]$ and $\channelPhaseUL[\indexED,\timeSymbol]$ as $\channelPhaseDLInitial[\indexED]+2\pi\timeSymbol {\velocity[\indexED]\cos(\angleOfArrival[\indexED])}/{\waveLength}
$ and $\channelPhaseULInitial[\indexED]+2\pi \timeSymbol {\velocity[\indexED]\cos(\angleOfArrival[\indexED])}/{\waveLength}
$, where $\channelPhaseDLInitial[\indexED]$ and $\channelPhaseULInitial[\indexED]$ are the random phase shifts in the DL and UL, respectively,  $\waveLength=\speedOfLigth/\carrierFrequency$ is the wavelength, $\speedOfLigth$ is the speed of light, and $\velocity[\indexED]$ and $\angleOfArrival[\indexED]$ denote the velocity for $\angleOfArrival[\indexED]\sim \uniformDistribution[0][2\pi]$ and the path angle relative to the BS location, respectively.

Suppose that $\numberOfOACSymbols$ parameters are transmitted over $\numberOfOACSymbols$ \ac{OFDM} symbols with the symbol duration $\symbolDurationOFDM$ and the \ac{CP} duration $\cpDuration$, where $\symbolEle[\indexED,\indexOACsymbol]$ is mapped to a fixed subcarrier of the $\indexOACsymbol$th \ac{OFDM} symbol, $\forall\indexOACsymbol$. The received symbol at the \ac{BS} can be expressed as\footnote{The amount of interference due to the \textit{residual} CFO is assumed to be negligible as it is significantly smaller than the subcarrier spacing in practice.
}
$\superposedSymbol[\indexOACsymbol] = \sum_{\indexED=1}^{\numberOfEdgeDevices} \receivedSymbolUL[\indexED][\timeSymbol]$,
where $\receivedSymbolUL[\indexED][\timeSymbol]$ is the received symbol from the $\indexED$th node at the \ac{BS}, and omitting the noise for clarity, it can be expressed as
\begin{align}
 \receivedSymbolUL[\indexED][\timeSymbol] = \channelAmpUL[\indexED,\timeSymbol]\constante^{\constantj\channelPhaseUL[\indexED,\timeSymbol]}\times \constante^{-\constantj(2\pi\CFO[\indexED]\timeSymbol+\PO[\indexED])}\times\symbolEle[\indexED,\indexOACsymbol],~\timeSymbol=\indexOACsymbol\symbolDurationOFDMTotal~,
\end{align}
for $\symbolDurationOFDMTotal\triangleq\symbolDurationOFDM+\cpDuration$. Thus, the phase of the received symbol of the $\indexED$th node is a function of the \ac{UL} channel, \ac{CFO}, and \ac{PO}, and rapidly changes over time as
\begin{align}
	\angle\receivedSymbolUL[\indexED][\timeSymbol] ={{\channelPhaseUL[\indexED,\timeSymbol]}-2\pi\CFO[\indexED]\timeSymbol-\PO[\indexED]}+\angle\symbolEle[\indexED,\indexOACsymbol]~.
	\label{eq:phaseOfUplinkSymbolFromNode}
\end{align}
The difficulty for \ac{OAC} arises from the fact that $\superposedSymbol[\indexOACsymbol]$ \textit{cannot} be expressed as
$\channelAmp[] \constante^{-\constantj(2\pi\CFO[]\timeSymbol+\PO[])}\times\sum_{\indexED=1}^{\numberOfEdgeDevices}\symbolEle[\indexED,\indexOACsymbol]$,
for some  parameters $\channelAmp[]$, $\CFO[]$, and $\PO[]$, \textit{independent} of the transmitted symbols, in general. Thus, typical estimation and correction methods (e.g., CFO estimation and correction, channel estimation, and linear equalization) relying on the existence of independent parameters like $\channelAmp[]$, $\CFO[]$, and $\PO[]$ are not helpful in obtaining the desired sum via receiver-only-based processing. 
Thus, to obtain the desired sum, the transmitters need to precode their transmission, which fundamentally requires phase synchronization in the network under the hardware imperfections and mobility in the environment.

Let $\transmittedSymbolDL[][]$ be a transmitted symbol from the BS in DL. The received symbol at the $\indexED$th node can be expressed as
$
	\receivedSymbolDL[\indexED][\timeSymbol]=\channelAmpDL[\indexED,\timeSymbol]\constante^{\constantj\channelPhaseDL[\indexED,\timeSymbol]}\times \constante^{\constantj(2\pi\CFO[\indexED]\timeSymbol+\PO[\indexED])}\times\transmittedSymbolDL[][]
$.
Hence, the phase of the received symbol at the $\indexED$th node is also a function of the \ac{DL} channel, CFO, and PO. Similar to $\angle\receivedSymbolUL[\indexED][\timeSymbol]$, $\angle\receivedSymbolDL[\indexED][\timeSymbol]$ also rapidly changes as a function of time as
$
	\angle\receivedSymbolDL[\indexED][\timeSymbol] ={{\channelPhaseDL[\indexED,\timeSymbol]}+2\pi\CFO[\indexED]\timeSymbol+\PO[\indexED]}+\angle\transmittedSymbolDL[][]
$, 
where the rotation due to the \ac{CFO} and \ac{PO} is in the reverse direction of the one in \ac{UL} channel (see \eqref{eq:phaseOfUplinkSymbolFromNode}).

\section{Phase-coded Pilots}
\label{sec:pcp}
\begin{figure*}
	\centering
	\includegraphics[width = 6.5in]{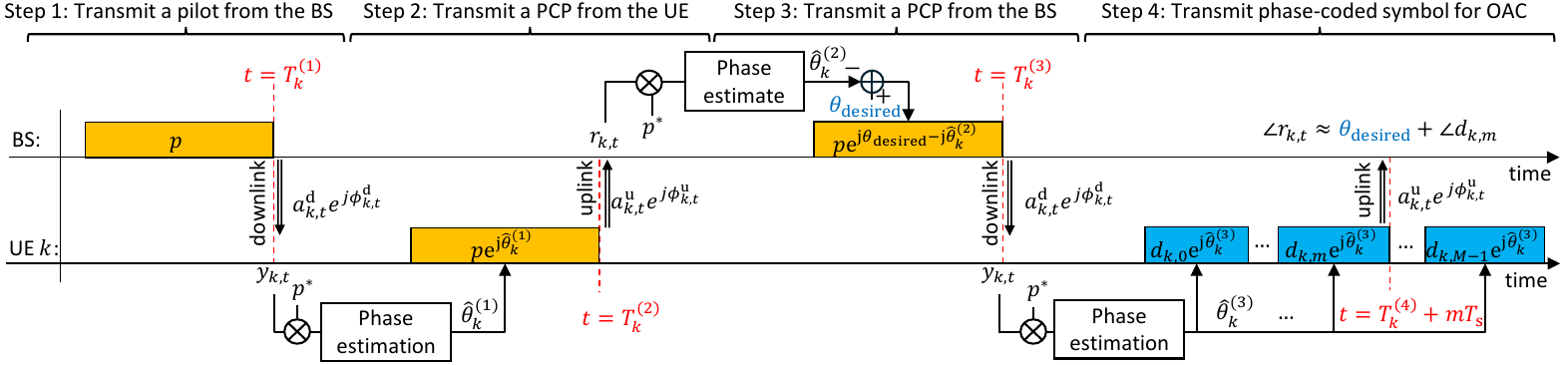}
	\caption{The basic steps of \acp{PCP} to achieve distributed phase synchronization in the network.}
	\label{fig:pcpwoAmp}
\end{figure*}
Suppose that the desired phase is $\desiredPhase$ for the received symbols from \textit{all} nodes at the \ac{BS} side. 
To align the phase of the received symbol from the $\indexED$th node to the desired angle $\desiredPhase$ at the \ac{BS} side,  the BS and the $\indexED$th node exchange \acp{PCP} to estimate and correct the round-trip phase change in the DL and UL. The steps of the proposed procedure, illustrated in \figurename~\ref{fig:pcpwoAmp}, is as follows:

\textbf{Step 1} (Request): The \ac{BS} transmits an uncoded pilot symbol $\pilotSymbol[]$. A pilot is typically a complex number for $|\pilotSymbol[]|=1$, e.g., a \ac{QPSK} symbol. Without loss of generality, we assume $\pilotSymbol[]=1$. The received symbol at the $\indexED$th node  can be expressed as
\begin{align}
	\receivedSymbolDL[\indexED][{\timeSymbol}]=\channelAmpDL[\indexED,\timeSymbol]\constante^{\constantj\channelPhaseDL[\indexED,\timeSymbol]}\times \constante^{\constantj(2\pi\CFO[\indexED]\timeSymbol+\PO[\indexED])}\times\pilotSymbol[]+\noiseSymbolStepOne[\indexED],~\timeSymbol=\instantStepOne[\indexED]~,
	\nonumber
\end{align}
where $\noiseSymbolStepOne[\indexED]$ is the noise at the $\indexED$th node.
By using the pilot, the $\indexED$th node estimates the phase change in the DL as 
\begin{align}
	\phaseEstimateStepOne[\indexED]&\triangleq\angle\pilotSymbol[]^*\receivedSymbolDL[\indexED][{\timeSymbol}]|_{\timeSymbol=\instantStepOne[\indexED]}=\channelPhaseDL[{\indexED,\instantStepOne[\indexED]}]+2\pi\CFO[\indexED]\instantStepOne[\indexED]+\PO[\indexED]+\noisePhaseAdditiveStepOne[\indexED]~,
	\nonumber
\end{align}
where $\noisePhaseAdditiveStepOne[\indexED]$ is the phase error due to the noise.

\textbf{Step 2} (Response): As a response, the $\indexED$th node  transmits a \ac{PCP}, i.e., $\pilotSymbol[]\constante^{\constantj\phaseEstimateStepOne[\indexED]}$, based on the estimated phase change $\phaseEstimateStepOne[\indexED]$ in Step~1. The received symbol at the BS and the estimated phase change in the UL channel can be written by
\begin{align}
	\receivedSymbolUL[\indexED][{\timeSymbol}]=\channelAmpUL[\indexED,\timeSymbol]\constante^{\constantj\channelPhaseUL[\indexED,\timeSymbol]}\times \constante^{-\constantj(2\pi\CFO[\indexED]\timeSymbol+\PO[\indexED])}\times\pilotSymbol[]\constante^{\constantj\phaseEstimateStepOne[\indexED]}+\noiseSymbolStepTwo[\indexED],~\timeSymbol=\instantStepTwo[\indexED]\nonumber
\end{align}
and $\phaseEstimateStepTwo[\indexED]\triangleq\angle\pilotSymbol[]^*\receivedSymbolUL[\indexED][{\timeSymbol}]|_{\timeSymbol=\instantStepTwo[\indexED]}$ is given by
\begin{align}
	&\phaseEstimateStepTwo[\indexED]=\phaseEstimateStepOne[\indexED]+ {\channelPhaseUL[{\indexED,\instantStepTwo[\indexED]}]}{-(2\pi\CFO[\indexED]\instantStepTwo[\indexED]+\PO[\indexED])}+\noisePhaseAdditiveStepTwo[\indexED]
	\label{eq:phaset1Change}\\&=2\pi\CFO[\indexED](\instantStepOne[\indexED]-\instantStepTwo[\indexED])+\channelPhaseDL[{\indexED,\instantStepOne[\indexED]}]+\channelPhaseUL[{\indexED,\instantStepTwo[\indexED]}]+\noisePhaseAdditiveStepOne[\indexED]+\noisePhaseAdditiveStepTwo[\indexED]~,\nonumber
\end{align}
respectively, where  $\noiseSymbolStepTwo[\indexED]$ is the noise and $\noisePhaseAdditiveStepTwo[\indexED]$ is the phase error due to the noise. As can be seen from \eqref{eq:phaset1Change}, the \ac{PO} $\PO[\indexED]$ is eliminated and the impact of \ac{CFO} on the phase is reduced as $2\pi\CFO[\indexED](\instantStepOne[\indexED]-\instantStepTwo[\indexED])\approx0$ for $\instantStepOne[\indexED]\approx\instantStepTwo[\indexED]$. The remaining phase in \eqref{eq:phaset1Change} is the round-trip phase change in DL and UL. 	

\textbf{Step 3} (Feedback): 
In this step,  the \ac{BS} provides feedback regarding the round-trip phase change and $\desiredPhase$ by transmitting a \ac{PCP} as $\pilotSymbol[]\constante^{\constantj(\desiredPhase-\phaseEstimateStepTwo[\indexED])}$. The received symbol at the $\indexED$th node for  $\timeSymbol=\instantStepThree[\indexED]$ can be expressed as
\begin{align}
	\receivedSymbolDL[\indexED][{\timeSymbol}]=\channelAmpDL[\indexED,\timeSymbol]\constante^{\constantj\channelPhaseDL[\indexED,\timeSymbol]}\times \constante^{\constantj(2\pi\CFO[\indexED]\timeSymbol+\PO[\indexED])}\times\pilotSymbol[]\constante^{\constantj(\desiredPhase-\phaseEstimateStepTwo[\indexED])}+\noiseSymbolStepThree[\indexED]~,\nonumber
\end{align}
where  $\noiseSymbolStepThree[\indexED]$ is the noise at the $\indexED$th node. The  node then estimates the phase change in the DL channel as 
\begin{align}
	\phaseEstimateStepThree[\indexED]&\triangleq\angle\pilotSymbol[]^*\receivedSymbolDL[\indexED][{\timeSymbol}]|_{\timeSymbol=\instantStepThree[\indexED]}\nonumber\\&=\desiredPhase-\phaseEstimateStepTwo[\indexED]+ {\channelPhaseDL[{\indexED,\instantStepThree[\indexED]}]}{+(2\pi\CFO[\indexED]\instantStepThree[\indexED]+\PO[\indexED])}+\noisePhaseAdditiveStepThree[\indexED]
	\nonumber\\&= \desiredPhase+2\pi\CFO[\indexED](\instantStepThree[\indexED]+\instantStepTwo[\indexED]-\instantStepOne[\indexED])+\PO[\indexED]\nonumber\\&~~~+\channelPhaseDL[{\indexED,\instantStepThree[\indexED]}]-\channelPhaseDL[{\indexED,\instantStepOne[\indexED]}]-\channelPhaseUL[{\indexED,\instantStepTwo[\indexED]}]+\noisePhaseAdditiveStepThree[\indexED]-\noisePhaseAdditiveStepTwo[\indexED]-\noisePhaseAdditiveStepOne[\indexED]~,\nonumber
\end{align}
where $\noisePhaseAdditiveStepThree[\indexED]$ is the phase error due to the noise.

\textbf{Step 4} (Aggregation):  Finally, for OAC, the $\indexED$th node  transmit the phase-coded parameter, i.e.,  $\symbolEle[\indexED,\indexOACsymbol]\constante^{\constantj\phaseEstimateStepThree[\indexED]}$, for coherent superposition. The $\indexOACsymbol$th received symbol can be written by $
	\superposedSymbol[\indexOACsymbol] = \sum_{\indexED=1}^{\numberOfEdgeDevices} \receivedSymbolUL[\indexED][\timeSymbol]+\noiseSymbolStepFour[\indexOACsymbol],~\timeSymbol=\instantStepFour[\indexED]+\indexOACsymbol\symbolDurationOFDMTotal$, where $\noiseSymbolStepFour[\indexOACsymbol]$ is the noise on the $\indexOACsymbol$th superposed symbol $\superposedSymbol[\indexOACsymbol] $ and $\receivedSymbolUL[\indexED][{\timeSymbol}]$ is given by
\begin{align}
	\receivedSymbolUL[\indexED][{\timeSymbol}]=&\channelAmpUL[\indexED,\timeSymbol]\constante^{\constantj\channelPhaseUL[\indexED,\timeSymbol]}\times \constante^{-\constantj(2\pi\CFO[\indexED]\timeSymbol+\PO[\indexED])}\times\symbolEle[\indexED,\indexOACsymbol]\constante^{\constantj\phaseEstimateStepThree[\indexED]}~.
\end{align}
Thus, the angle of $\receivedSymbolUL[\indexED][{\timeSymbol}]$ can be expressed as
\begin{align}
	\phaseEstimateStepFour[\indexED,\indexOACsymbol]&\triangleq\angle\receivedSymbolUL[\indexED][{\timeSymbol}]|_{\timeSymbol=\instantStepFour[\indexED]+\indexOACsymbol\symbolDurationOFDMTotal}\nonumber\\&=\phaseEstimateStepThree[\indexED]+ {\channelPhaseUL[{\indexED,\instantStepFour[\indexED]+\indexOACsymbol\symbolDurationOFDMTotal}]}{-(2\pi\CFO[\indexED](\instantStepFour[\indexED]+\indexOACsymbol\symbolDurationOFDMTotal)+\PO[\indexED])}
	\nonumber\\&=\desiredPhase+\angle\symbolEle[\indexED,\indexOACsymbol]+\errPhaseCFO[\indexED,\indexOACsymbol] + \errPhaseMobility[\indexED,\indexOACsymbol] + 	\errPhaseNoise[\indexED,\indexOACsymbol]~,
\end{align}
for
\begin{align}
	\errPhaseCFO[\indexED,\indexOACsymbol]&\triangleq2\pi\CFO[\indexED](\instantStepThree[\indexED]-\instantStepFour[\indexED]+\instantStepTwo[\indexED]-\instantStepOne[\indexED]+\indexOACsymbol\symbolDurationOFDMTotal)~,\label{eq:errPhaseCFO}
\end{align}
\begin{align}
	\errPhaseMobility[\indexED,\indexOACsymbol]&\triangleq\channelPhaseDL[{\indexED,\instantStepFour[\indexED]+\indexOACsymbol\symbolDurationOFDMTotal}]-\channelPhaseDL[{\indexED,\instantStepTwo[\indexED]}]+\channelPhaseUL[{\indexED,\instantStepThree[\indexED]}]-\channelPhaseUL[{\indexED,\instantStepOne[\indexED]}] \label{eq:errPhaseCHN} \\ &= 2\pi \frac{\velocity[\indexED]\cos(\angleOfArrival[\indexED])(\instantStepThree[\indexED]-\instantStepOne[\indexED]+\instantStepFour[\indexED]-\instantStepTwo[\indexED]+\indexOACsymbol\symbolDurationOFDMTotal)}{\waveLength}~,\nonumber
\end{align}
and $\errPhaseNoise[\indexED,\indexOACsymbol]\triangleq\noisePhaseAdditiveStepThree[\indexED]-\noisePhaseAdditiveStepTwo[\indexED]-\noisePhaseAdditiveStepOne[\indexED]$. Hence,  $\angle\receivedSymbolUL[\indexED][{\timeSymbol}]$ in the last step is approximately equal to $\desiredPhase+\angle\symbolEle[\indexED,\indexOACsymbol]$ for $\errPhaseCFO[\indexED,\indexOACsymbol] + \errPhaseMobility[\indexED,\indexOACsymbol] + 	\errPhaseNoise[\indexED,\indexOACsymbol]\approx0$~radians, where  the phase error is a function of the amount of residual CFO, the timing of the exchanged packets, the mobility in the environment,   and the noise at the radios. 

\subsection{CFO-Resilient Procedure for Multiple Nodes}
\label{subsec:CFOresilientProcedure}
\begin{figure*}
	\centering
	\includegraphics[width = 7in]{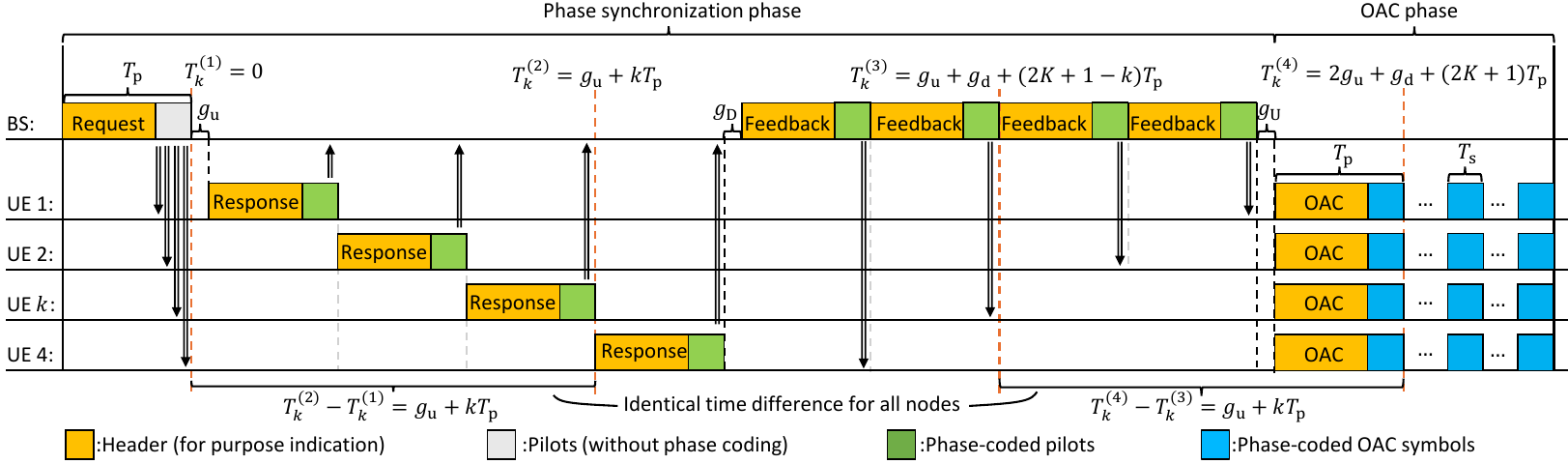}
	\caption{CFO-resilient multi-user protocol with \acp{PCP}. Since $\instantStepFour[\indexED]-\instantStepThree[\indexED]=\instantStepTwo[\indexED]-\instantStepOne[\indexED]$ holds for \textit{all} nodes, $\errPhaseCFO[\indexED,\indexOACsymbol]$ is not a function of number of nodes.}
	\label{fig:muprotocol}
\end{figure*}
One of the key observations from \eqref{eq:errPhaseCFO} is that the impact of the residual \ac{CFO} of the phase synchronization can be mitigated considerably for $\instantStepFour[\indexED]-\instantStepThree[\indexED]\approx\instantStepTwo[\indexED]-\instantStepOne[\indexED]$.  To exploit this key property, we propose a procedure for multiple nodes, where the order of the nodes are reversed during the UL and DL transmissions such that $\instantStepFour[\indexED]-\instantStepThree[\indexED]=\instantStepTwo[\indexED]-\instantStepOne[\indexED]$ holds for \textit{all} nodes. To elaborate the proposed method, let $\packetDuration$ be the packet duration for each aforementioned step.

 As shown in \figurename~\ref{fig:muprotocol}, before the \ac{OAC} phase, a phase-synchronization phase is initiated by the BS by transmitting a request packet including a pilot without any phase coding (i.e., Step 1), common for all $\numberOfEdgeDevices$ nodes. Once the nodes receive the request packet (i.e., $\instantStepOne[\indexED]=0$), each node access the channel in an order 
  and transmits a response packet in the UL, including a \ac{PCP} (i.e., Step~2).  Let $\guardTimeUplink$ be the minimum response time of a node after it receives a packet. Thus,  the instant of the received response packet of the $\indexED$th node at the BS can be expressed as $\instantStepTwo[\indexED]=\guardTimeUplink+\indexED\packetDuration$. The BS then transmit feedback packets, including \acp{PCP} (i.e., Step~3), for the all nodes in the reverse order. For instance, the first feedback symbol is for the node transmitting the last response packet  in the UL. Similarly, the last feedback symbol is for the  node transmitting the first response packet, as can be seen in \figurename~\ref{fig:muprotocol}. Let $\guardTimeDownlink$ be the response time of the BS. We can then express the instant of the received feedback symbol at the $\indexED$th node as $\instantStepThree[\indexED]=\guardTimeUplink+\guardTimeDownlink+(2\numberOfEdgeDevices+1-\indexED)\packetDuration$.

 After all the nodes receive the feedback symbols, the nodes transmit the phase-coded OAC symbols, simultaneously, during the OAC phase, and the OAC symbol is received at  $\instantStepFour[\indexED]=2\guardTimeUplink+\guardTimeDownlink+(2\numberOfEdgeDevices+1)\packetDuration$. Thus, the proposed procedure maintains an identical time difference for all nodes, i.e., $
 	\instantStepTwo[\indexED]-\instantStepOne[\indexED]=\instantStepFour[\indexED]-\instantStepThree[\indexED]=\guardTimeUplink+\indexED\packetDuration~,\forall\indexED$,  eliminating the impact of the residual CFOs of all nodes on the first OAC symbol. With the proposed procedure, \eqref{eq:errPhaseCFO} can be re-expressed as
$	\errPhaseCFO[\indexED,\indexOACsymbol] = 2\pi\CFO[\indexED]\indexOACsymbol\symbolDurationOFDMTotal
 	$.
Now, suppose that $\CFO[\indexED]\sim\gaussian[\varianceCFO][0]$ holds for all nodes, where $\varianceCFO$ is the variance of the residual CFO in the network. Hence, $\errPhaseCFO[\indexED,\indexOACsymbol] $  can be modeled as a zero-mean Gaussian random variable with
\begin{align}
	\expectationOperator[|{\errPhaseCFO[\indexED,\indexOACsymbol]}|^2][]=4\pi^2\indexOACsymbol^2\symbolDurationOFDMTotal^2\varianceCFO&~.
	\label{eq:mseCFO}
\end{align}
Note that $\errPhaseCFO[\indexED,\indexOACsymbol]$ and $\expectationOperator[|{\errPhaseCFO[\indexED,\indexOACsymbol]}|^2][]$ do \textit{not} depend on the duration of the phase-synchronization phase with the proposed procedure, therefore, the number of nodes in the network.

\subsubsection{The Impact of Noise on Phase Synchronization}

To analyze the impact of the noise on phase synchronization, i.e., $\noisePhaseAdditiveStepOne[\indexED]$, $\noisePhaseAdditiveStepTwo[\indexED]$, and $\noisePhaseAdditiveStepThree[\indexED]$, let us consider a generic expression  as $\receivedSymbol=\channelAmp[]\constante^{\constantj\phaseSymbolWithoutNoise[]}+\noiseSymbol[]$ for  $\noiseSymbol[]=\noiseAmp[]\constante^{\constantj\noisePha[]}$. We can then express the angle of $\receivedSymbol$ as $\phaseSymbolWithoutNoise[]+	\noisePhase[]$ for
$
	\noisePhase[]= \tan^{-1}\left( \frac{\noiseAmp[]\sin(\noisePha[]-\phaseSymbolWithoutNoise[])}{\channelAmp[]+\noiseAmp[]\cos(\noisePha[]-\phaseSymbolWithoutNoise[])}\right)\approx\frac{\noiseAmp[]}{\channelAmp[\indexED]}\sin(\noisePha[]-\phaseSymbolWithoutNoise[])
$,
 where the approximation holds for $\noiseAmp[]\ll\channelAmp[]$. As the argument of the sine function is uniformly distributed,  $\noisePhase[]$ can be modeled as a zero-mean random variable with the variance of  $\noiseVariance/(2\channelAmp[\indexED])$  for $\noiseSymbol[]\sim\complexGaussian[\noiseVariance][0]$ and $\noiseAmp[]\ll\channelAmp[]$. Therefore, by assuming independent random variables $\noiseSymbolStepOne[\indexED]$, $\noiseSymbolStepTwo[\indexED]$, and $\noiseSymbolStepThree[\indexED]$ for $\noiseSymbolStepOne[\indexED],\noiseSymbolStepThree[\indexED]\sim\complexGaussian[\noiseVarianceES][0]$ and $\noiseSymbolStepTwo[\indexED]\sim\complexGaussian[\noiseVarianceED][0]$, the phase deviation due to the noise, i.e., $\errPhaseNoise[\indexED,\indexOACsymbol]$, can be approximated as a zero-mean Gaussian random variable with
\begin{align}
	\expectationOperator[|{\errPhaseNoise[\indexED,\indexOACsymbol]}|^2][]=&\frac{\noiseVarianceES}{2\channelAmpUL[{\indexED,\instantStepTwo[\indexED]}]}+\frac{\noiseVarianceED}{2\channelAmpDL[{\indexED,\instantStepThree[\indexED]}]}+\frac{\noiseVarianceED}{2\channelAmpDL[{\indexED,\instantStepOne[\indexED]}]}~.
	\label{eq:mseNoise}
\end{align}

\subsubsection{The Impact of Mobility on Phase Synchronization}
For the proposed procedure,  \eqref{eq:errPhaseCHN} can be re-expressed as
\begin{align}
	\errPhaseMobility[\indexED,\indexOACsymbol] &= \frac{2\pi\velocity[\indexED]\cos(\angleOfArrival[\indexED])}{\waveLength} (2(\guardTimeUplink+\guardTimeDownlink+(2\numberOfEdgeDevices+1-\indexED)\packetDuration)+\indexOACsymbol\symbolDurationOFDMTotal)~.\nonumber
\end{align}
For the velocity distribution, one may consider two models. For the first model, $\velocity[\indexED]=\maxVelocity$ is fixed for all nodes. In this case, $\errPhaseMobility[\indexED,\indexOACsymbol]$ is a zero-mean random variable following the Jakes spectrum with
\begin{align}
	\expectationOperator[|{\errPhaseMobility[\indexED,\indexOACsymbol]}|^2][]=\frac{2\pi^2\maxVelocity^2}{\waveLength^2}  (2(\guardTimeUplink+\guardTimeDownlink+(2\numberOfEdgeDevices+1-\indexED)\packetDuration)+\indexOACsymbol\symbolDurationOFDMTotal)^2,
	\label{eq:mseMob}
\end{align}
For the second model,  $\velocity[\indexED]$ is a random variable with Rayleigh distribution with the mean $\velocityMean$, leading to  a zero-mean Gaussian $\errPhaseMobility[\indexED,\indexOACsymbol]$ with
\begin{align}
	\expectationOperator[|{\errPhaseMobility[\indexED,\indexOACsymbol]}|^2][]=\frac{8\pi\velocityMean^2}{\waveLength^2}  (2(\guardTimeUplink+\guardTimeDownlink+(2\numberOfEdgeDevices+1-\indexED)\packetDuration)+\indexOACsymbol\symbolDurationOFDMTotal)^2~,
	\label{eq:mseMobGaussian}
\end{align}
The second model's advantage is that the distribution of phase deviations can be analyzed analytically, discussed next.


\subsubsection{Phase Deviation Distribution}
Let $\angle\symbolEle[\indexED,\indexOACsymbol]$ be $0$. Hence, by using \eqref{eq:mseCFO}, \eqref{eq:mseNoise}, and \eqref{eq:mseMob} (or \eqref{eq:mseMobGaussian}) we can express the \ac{RMSE} of  $\phaseEstimateStepFour[\indexED,\indexOACsymbol]$ for the $\indexED$th node as
\begin{align}
	&\RMSE[\indexED]\{\phaseEstimateStepFour[\indexED,\indexOACsymbol]\}={\expectationOperator[|	\phaseEstimateStepFour[\indexED,\indexOACsymbol] - \desiredPhase|^2][]}^{\frac{1}{2}}=\stdPhaseError[\indexED][\indexOACsymbol]~.
	\label{eq:rmseResultTheory}
\end{align}
where $\stdPhaseError[\indexED][\indexOACsymbol]\triangleq(\expectationOperator[|{\errPhaseNoise[\indexED,\indexOACsymbol]}|^2][] + 
\expectationOperator[|{\errPhaseMobility[\indexED,\indexOACsymbol]}|^2][] +	\expectationOperator[|{\errPhaseCFO[\indexED,\indexOACsymbol]}|^2][])^\frac{1}{2}$ is the standard deviation of the phase errors for the $\indexED$th node and $\indexOACsymbol$th \ac{OAC} symbol.
Since $\errPhaseNoise[\indexED,\indexOACsymbol]$, $\errPhaseCFO[\indexED,\indexOACsymbol]$, and $\errPhaseMobility[\indexED,\indexOACsymbol]$ can be modeled as zero-mean Gaussian random variables with the variances given in \eqref{eq:mseCFO}, \eqref{eq:mseNoise},  and \eqref{eq:mseMobGaussian}, respectively, finally, the \ac{CDF} of the absolute phase deviations can  be calculated as
\begin{align}
\probability[{|\phaseEstimateStepFour[\indexED,\indexOACsymbol]|\le\theta}]=2\normalCDF[\frac{\theta}{\stdPhaseError[\indexED][\indexOACsymbol]}]-1~.
\label{eq:cdfPhase}
\end{align}

\subsubsection{Computation Rate}
The proposed procedure consumes $2\guardTimeUplink+\guardTimeDownlink+2(\numberOfEdgeDevices+1)\packetDuration+(\numberOfOACSymbols-1)\symbolDurationOFDMTotal$~seconds
to compute $\numberOfOACSymbols$ functions over a single \ac{OFDM} subcarrier. Hence, its computation rate is
${\numberOfOACSymbols\symbolDurationOFDM}/({2\guardTimeUplink+\guardTimeDownlink+2(\numberOfEdgeDevices+1)\packetDuration+(\numberOfOACSymbols-1)\symbolDurationOFDMTotal})$ functions/(s$\cdot$Hz). Note that the computation rate of the traditional first-communicate-then-compute approach can  be  approximated as ${\spectralEfficiency}/({\numberOfBitsForQuantization\numberOfEdgeDevices})$~functions/(s$\cdot$Hz), where $\spectralEfficiency$ is the spectral efficiency in bits/(s$\cdot$Hz) and $\numberOfBitsForQuantization$ is the number of bits for parameter quantization.

\section{Numerical Results}
\label{sec:numerical}
In this section, we assess  the proposed procedure numerically for $\numberOfEdgeDevices\in\{5,10,20\}$~nodes, $\maxVelocity=\{0.1, 1.5\}$~m/s (for \eqref{eq:mseMob}), $\velocityMean=\{0.1, 1.5\}$~m/s (for \eqref{eq:mseMobGaussian}), $\varianceCFO\in\{100,1000\}$, $\carrierFrequency=1.8$~GHz, $\guardTimeDownlink=\guardTimeUplink=16$~$\mu$s, $\symbolDurationOFDM=1/60\text{e}3$~$\mu$s, and $\cpDuration=\symbolDurationOFDM/8$. We assume an implementation of the proposed protocol with $\packetDuration=\symbolDurationOFDMTotal=18.75$~$\mu$s. We set $\desiredPhase=0$~radians, $\channelAmpDL[\indexED,\timeSymbol]=\channelAmpUL[\indexED,\timeSymbol]=1$, $\SNRatED\triangleq1/\noiseVarianceED\in\{20,30\}$~dB and $\SNRatES\triangleq1/\noiseVarianceES\in\{10,30\}$~dB, assuming a power-control loop in the network.

\begin{figure}
	\centering
	\includegraphics[width = \figuresize]{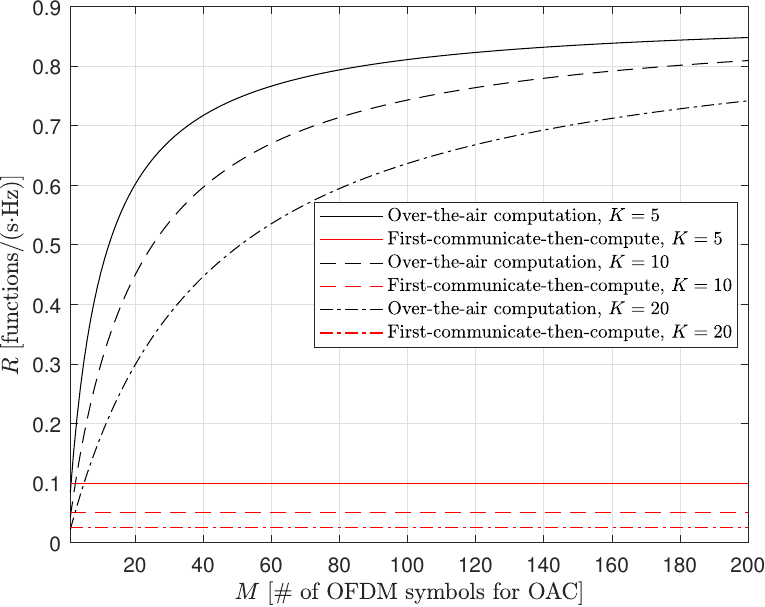}
	\caption{The computation rate versus the number of OFDM symbols for OAC with the proposed procedure.}
	\label{fig:computationRate}
\end{figure}
In \figurename~\ref{fig:computationRate}, we first compare the computation rates with the proposed procedure and the traditional first-communicate-then-compute approach for $\numberOfBitsForQuantization=8$~bits and $\spectralEfficiency=4$ bits/(s$\cdot$Hz). As can be seen from the figure, the computation rate with OAC is notably larger than the one with the traditional approach, and the efficiency of the proposed procedure improves with the number of OFDM symbols for OAC. We next analyze how many OFDM symbols can be used during the OAC phase under mobility and hardware impairments. 

\begin{figure}
	\centering
	\subfloat[$\numberOfEdgeDevices=5$~nodes.]{\includegraphics[width = \figuresize]{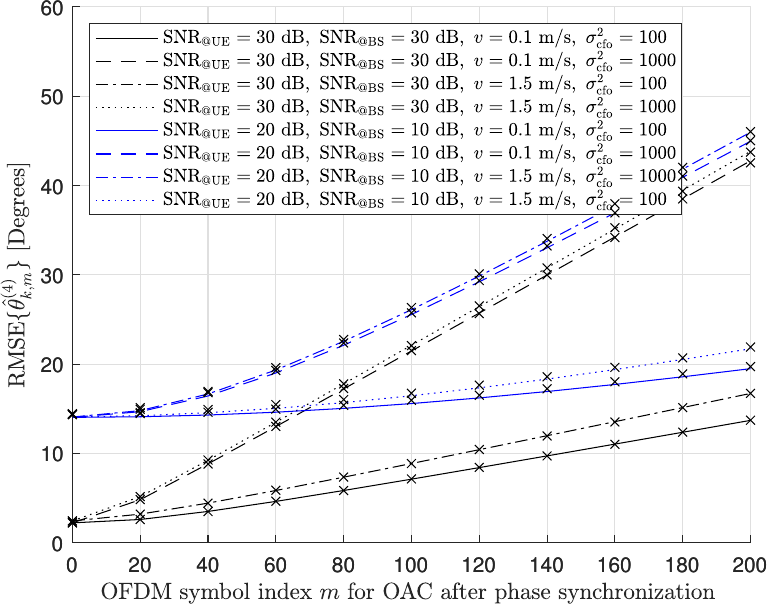}\label{subfig:rmse_5K}}\\
	\subfloat[$\numberOfEdgeDevices=20$~nodes.]{\includegraphics[width = \figuresize]{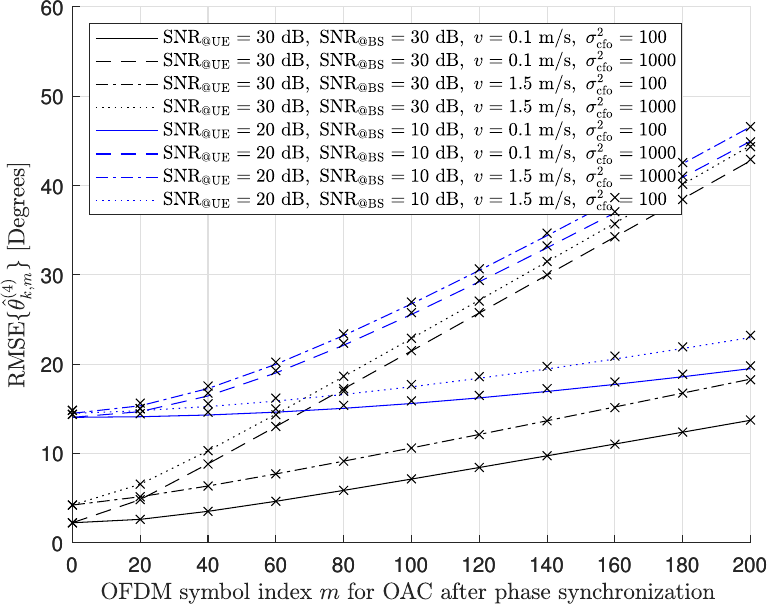}\label{subfig:rmse_20K}}
	\caption{The RMSE of phase deviations as a function of the OFDM symbol index for the worst node (i.e., $\indexED=1$)  ($\times$: simulation, lines: theory \eqref{eq:rmseResultTheory}).}
	\label{fig:phaseDeviation}
\end{figure}

In \figurename~\ref{fig:phaseDeviation}, we assess how the \ac{RMSE} gradually changes during the OAC phase under various mobility, \ac{SNR}, and impairment configurations for the worst node, i.e., $\indexED=1$, for $\numberOfEdgeDevices\in\{5,20\}$. Our first observation is that the amount of the noise at the nodes and  \ac{BS} jointly define the minimum \ac{RMSE}. While the minimum RMSE is about $3^\circ$ for the high-SNR configurations ($\SNRatED=\SNRatES=30$~dB), it increases to $15^\circ$ for the low-SNR configurations ($\SNRatED=20$~dB, $\SNRatES=10$~dB). Secondly, the RMSE gradually increases with $\indexOACsymbol$, limiting the number of OFDM symbols in the OAC phase under higher mobility and/or larger CFO variations. For example, assuming that $20^\circ$ is the maximum tolerable RMSE,  $\numberOfOACSymbols=80$ OFDM symbols can be used during the OAC phase for $\varianceCFO=1000$, dropping from $200$ OFDM symbols for $\varianceCFO=100$. Our third observation is that the impact of residual CFO on the phase deviations is more noticeable than the one for the mobility in the channel. For instance, if the mobility increases from $0.1$~m/s to $1.5$~m/s for $\varianceCFO=100$, the RMSE increases slightly. However, if $\varianceCFO$ is increased from $100$  to $1000$  for low mobility, the RMSE increases substantially, which indicates the need for an accurate CFO pre-compensation for reliable OAC. Finally, increasing the number of nodes from $5$ to $20$ has a non-negligible effect on the phase deviations as the duration of the phase synchronization phase increases. Hence, the efficiency of phase synchronization phase needs to be improved to support a large number of nodes. To this end, one effective solution is to parallelize the users within the coherence bandwidth during the phase-synchronization phase.  Note that multiple subcarriers within the coherence bandwidth can also be exploited to improve $\SNRatED$ and $\SNRatES$.

\begin{figure}
	\centering
	\subfloat[The distribution for the $\indexOACsymbol=20$th OFDM symbol ($\numberOfEdgeDevices=20$~nodes).]{\includegraphics[width=\figuresize]{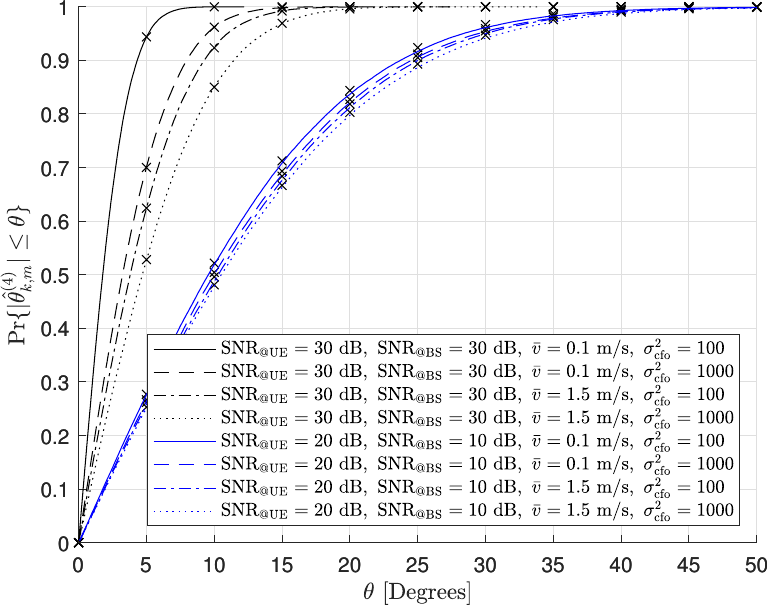}\label{subfig:cdf20m_20K}}	
	\\
	\subfloat[The distribution for the $\indexOACsymbol=100$th OFDM symbol  ($\numberOfEdgeDevices=20$~nodes).]{\includegraphics[width=\figuresize]{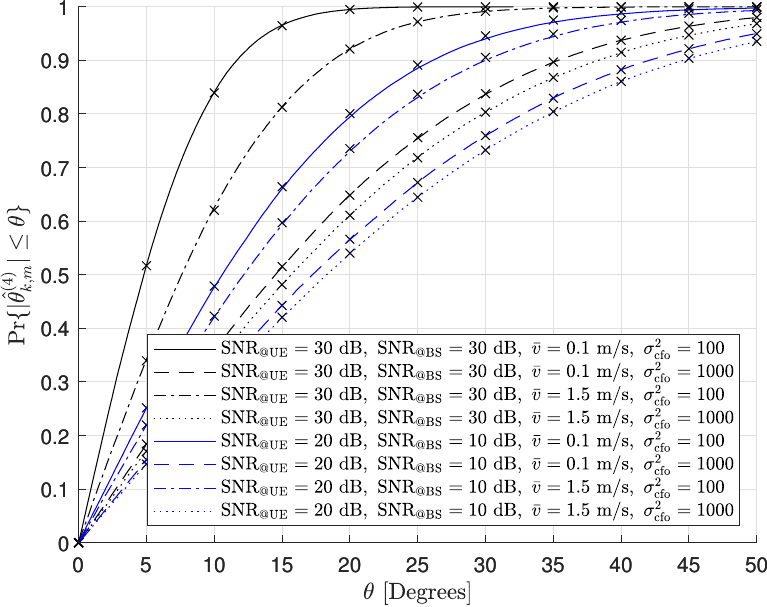}\label{subfig:cdf100m_20K}}	
	\caption{The CDF of the absolute phase deviations  ($\times$: theory \eqref{eq:cdfPhase}, lines: simulation). The residual CFO degrades the phase coherency considerably.}
	\label{fig:statisticsPhase}
\end{figure}
In \figurename~\ref{fig:statisticsPhase}, we analyze the distribution of the phase deviations for $\indexOACsymbol\in\{20,100\}$ for $\numberOfEdgeDevices=20$ nodes. In \figurename~\ref{fig:statisticsPhase}\subref{subfig:cdf20m_20K}, for all high-SNR configurations, $\probability[{|\phaseEstimateStepFour[\indexED,\indexOACsymbol]|>15^\circ}]$  is 5\%, while it increases to 30\% for all low-SNR configurations. 
In \figurename~\ref{fig:statisticsPhase}\subref{subfig:cdf100m_20K}, we perform the same analysis for the $100$th OFDM symbol. For the scenario with high SNR,  large CFO variation, and high mobility, the probability increases  to 50\%  for $\indexOACsymbol=100$th OFDM symbol, indicating a major limitation in the duration of OAC phase. We also obverse that residual CFO degrades the phase coherency more quickly than the mobility.

\begin{figure}
	\centering
	\includegraphics[width = 3in]{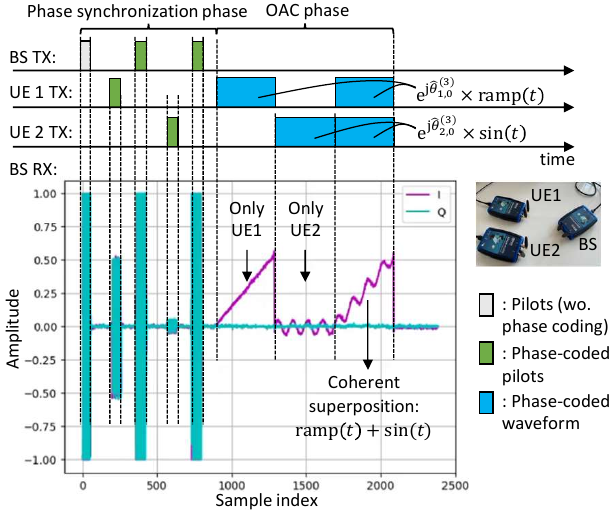}
	\caption{Proof-of-concept demonstration for coherent signal superposition with PCPs  with off-the-shelf SDRs. The BS's signal saturates at its TX and RX antennas are adjacent to each other.}
	\label{fig:demo}
\end{figure}
Finally, in \figurename~\ref{fig:demo}, we provide a proof-of-concept demonstration by using Adalm Pluto \acp{SDR}. In this demo, the SDRs are triggered to transmit a phase-coded ramp  and a sine function  simultaneously as  test waveforms, respectively, to demonstrate the coherent superposition.
To this end, we modify the \ac{FPGA} of the SDRs such that the radios estimate the phases in the UL and DL directions.  We do not use \textit{any} auxiliary synchronization method (e.g., GPS) for the plausibility of demo. In  \figurename~\ref{fig:demo}, we show the IQ data samples at the BS while the BS, UE1, and UE2 transmit. \figurename~\ref{fig:demo} shows that  the phase of each signal is well-aligned at $0^\circ$ (i.e., the samples at quadrature are almost zero-valued) and the signals coherently superpose, which demonstrates the feasibility of phase synchronization with PCPs. We refer the reader to our 802.11 AI/ML standard contribution in \cite{sahin_2025ieee} for further details due to the page limitations in this study.

\section{Concluding Remarks}
\label{sec:conc}
In this study, we  analyze the phase synchronization in a network for \ac{OAC}. With the proposed approach, the nodes and BS exchange \acp{PCP} such that the OAC symbols are aligned at a desired phase at the BS during the OAC phase by eliminating the round-trip phase change. We theoretically analyze the phase deviations and derive the RMSE and statistics of the phase deviation as a function of the OFDM symbol index for a given set of parameters such as residual CFO, velocity, the SNR in the UL and DL directions, number of nodes, and packet duration. We also provide proof-of-concept demonstration by using off-the-shelf SDRs to show the feasibility of the PCPs. Our results indicate that residual CFO is a primary limitation for maintaining phase-coherency in the network and gradually destroys the coherency over time more quickly than the mobility in the channel. The extension of this work will provide more details on hardware implementation and analysis regarding amplitude alignment.



\section*{Acknowledgment}
The author would like to thank Rui Yang and his team at InterDigital for the comprehensive discussions on PCPs.

\bibliographystyle{IEEEtran}
\bibliography{references}

\begin{thebibliography}{10}
\providecommand{\url}[1]{#1}
\csname url@samestyle\endcsname
\providecommand{\newblock}{\relax}
\providecommand{\bibinfo}[2]{#2}
\providecommand{\BIBentrySTDinterwordspacing}{\spaceskip=0pt\relax}
\providecommand{\BIBentryALTinterwordstretchfactor}{4}
\providecommand{\BIBentryALTinterwordspacing}{\spaceskip=\fontdimen2\font plus
\BIBentryALTinterwordstretchfactor\fontdimen3\font minus
  \fontdimen4\font\relax}
\providecommand{\BIBforeignlanguage}[2]{{%
\expandafter\ifx\csname l@#1\endcsname\relax
\typeout{** WARNING: IEEEtran.bst: No hyphenation pattern has been}%
\typeout{** loaded for the language `#1'. Using the pattern for}%
\typeout{** the default language instead.}%
\else
\language=\csname l@#1\endcsname
\fi
#2}}
\providecommand{\BIBdecl}{\relax}
\BIBdecl

\bibitem{Jeon_2014}
S.-W. Jeon, C.-Y. Wang, and M.~Gastpar, ``Computation over {Gaussian} networks
  with orthogonal components,'' \emph{IEEE Transactions on Information Theory},
  vol.~60, no.~12, pp. 7841--7861, 2014.

\bibitem{sahinSurvey2023}
A.~\c{S}ahin and R.~Yang, ``A survey on over-the-air computation,'' \emph{IEEE
  Commun. Surveys Tuts.}, vol.~25, no.~3, pp. 1877--1908, Apr. 2023.

\bibitem{Zhibin_2022oac}
Z.~Wang, Y.~Zhao, Y.~Zhou, Y.~Shi, C.~Jiang, and K.~B. Letaief, ``Over-the-air
  computation for {6G}: Foundations, technologies, and applications,''
  \emph{IEEE Internet of Things Journal}, pp. 1--25, 2024.

\bibitem{pérezneira2024waveformscomputingair}
\BIBentryALTinterwordspacing
A.~Pérez-Neira, M.~Martinez-Gost, A.~Şahin, S.~Razavikia, C.~Fischione, and
  K.~Huang, ``Waveforms for computing over the air,'' 2024. [Online].
  Available: \url{https://arxiv.org/abs/2405.17007}
\BIBentrySTDinterwordspacing

\bibitem{Xiaowen_2022}
X.~Cao, Z.~Lyu, G.~Zhu, J.~Xu, L.~Xu, and S.~Cui, ``An overview on over-the-air
  federated edge learning,'' \emph{IEEE Wireless Communications}, vol.~31,
  no.~3, pp. 202--210, 2024.

\bibitem{Cai_2018}
S.~Cai and V.~K.~N. Lau, ``Modulation-free {M2M} communications for
  mission-critical applications,'' \emph{IEEE Transactions on Signal and
  Information Processing over Networks}, vol.~4, no.~2, pp. 248--263, 2018.

\bibitem{Xiugang_2016}
X.~Wu, S.~Zhang, and A.~Özgür, ``{STAC}: Simultaneous transmitting and air
  computing in wireless data center networks,'' \emph{IEEE J. Sel. Areas
  Commun.}, vol.~34, no.~12, pp. 4024--4034, 2016.

\bibitem{Guo_2021}
H.~Guo, Y.~Zhu, H.~Ma, V.~K.~N. Lau, K.~Huang, X.~Li, H.~Nong, and M.~Zhou,
  ``Over-the-air aggregation for federated learning: Waveform superposition and
  prototype validation,'' \emph{Journal of Communications and Information
  Networks}, vol.~6, no.~4, pp. 429--442, 2021.

\bibitem{Lizhao_2024tmc}
L.~You, X.~Zhao, R.~Cao, Y.~Shao, and L.~Fu, ``Broadband digital over-the-air
  computation for wireless federated edge learning,'' \emph{IEEE Trans. Mobile
  Comput.}, vol.~23, no.~5, pp. 5212--5228, 2024.

\bibitem{Moghadam2014}
N.~N. Moghadam, H.~Farhadi, P.~Zetterberg, M.~N. Khormuji, and M.~Skoglund,
  ``Interference alignment — practical challenges and test-bed
  implementation,'' in \emph{Contemp. Issues in Wireless Commun.}, 2014.

\bibitem{Mudumbai_2007distBeaming}
R.~Mudumbai, G.~Barriac, and U.~Madhow, ``On the feasibility of distributed
  beamforming in wireless networks,'' \emph{IEEE Transactions on Wireless
  Communications}, vol.~6, no.~5, pp. 1754--1763, 2007.

\bibitem{Mghabghab_2021}
S.~R. Mghabghab and J.~A. Nanzer, ``Open-loop distributed beamforming using
  wireless frequency synchronization,'' \emph{IEEE Transactions on Microwave
  Theory and Techniques}, vol.~69, no.~1, pp. 896--905, 2021.

\bibitem{Sachin_2008}
S.~Katti, H.~Rahul, W.~Hu, D.~Katabi, M.~Medard, and J.~Crowcroft, ``{XORs} in
  the air: Practical wireless network coding,'' \emph{IEEE/ACM Transactions on
  Networking}, vol.~16, no.~3, pp. 497--510, 2008.

\bibitem{sahin_2025ieee}
A.~\c{S}ahin and et~al., ``Feasibility study of phase-synchronization for
  wireless federated learning on {WLAN},'' {IEEE} 802.11-25/304r0, Mar. 2025.

\end{thebibliography}

\end{document}